\documentclass[conference, 10pt, nofonttune]{IEEEtran}
\IEEEoverridecommandlockouts

\usepackage{cite}
\usepackage{amsmath,amssymb,amsfonts}
\usepackage{graphicx}
\usepackage{booktabs}
\usepackage{array}
\usepackage{xcolor}
\usepackage{url} 
\usepackage{textcomp}
\usepackage{tikz}
\usetikzlibrary{positioning,fit,arrows.meta,calc,decorations.pathmorphing}
\usepackage{pgfplots}
\pgfplotsset{compat=1.18}
\usepgfplotslibrary{statistics}

\usepackage[acronym,nomain]{glossaries}
\glsdisablehyper
\newacronym{aclr}{ACLR}{Adjacent-Channel Leakage Ratio}
\newacronym{ai}{AI}{artificial intelligence}
\newacronym{amf}{AMF}{Access and Mobility Management Function}
\newacronym{api}{API}{application programming interface}
\newacronym{arfcn}{ARFCN}{Absolute Radio Frequency Channel Number}
\newacronym{bler}{BLER}{Block Error Rate}
\newacronym{bpf}{BPF}{bandpass filter}
\newacronym{cbrs}{CBRS}{Citizens Broadband Radio Service}
\newacronym{cdf}{CDF}{cumulative distribution function}
\newacronym{cdqn}{C-DQN}{Constrained Double Deep Q-Network}
\newacronym{cept}{CEPT}{European Conference of Postal and Telecommunications Administrations}
\newacronym{cmdp}{CMDP}{Constrained Markov Decision Process}
\newacronym{cn5g}{5GC}{5G Core}
\newacronym{cots}{COTS}{commercial off-the-shelf}
\newacronym{cp}{CP}{cyclic prefix}
\newacronym{cpo}{CPO}{Constrained Policy Optimization}
\newacronym{cpu}{CPU}{central processing unit}
\newacronym{cqi}{CQI}{Channel Quality Indicator}
\newacronym{ddpg}{DDPG}{Deep Deterministic Policy Gradient}
\newacronym{dl}{DL}{Downlink}
\newacronym{dlsch}{DL-SCH}{Downlink Shared Channel}
\newacronym{dpdk}{DPDK}{Data Plane Development Kit}
\newacronym{dqn}{DQN}{Deep Q-Network}
\newacronym{drb}{DRB}{Data Radio Bearer}
\newacronym{dsa}{DSA}{dynamic spectrum allocation}
\newacronym{dss}{DSS}{dynamic spectrum sharing}
\newacronym{e2smrc}{E2SM-RC}{E2 Service Model for RAN Control}
\newacronym{ecpri}{eCPRI}{enhanced common public radio interface}
\newacronym{eess}{EESS}{Earth Exploration Satellite Service}
\newacronym{eirp}{EIRP}{Effective Isotropic Radiated Power}
\newacronym{embb}{eMBB}{enhanced Mobile Broadband}
\newacronym{esc}{ESC}{Environmental Sensing Capability}
\newacronym{evm}{EVM}{Error Vector Magnitude}
\newacronym{fft}{FFT}{Fast Fourier Transform}
\newacronym{fr1}{FR1}{Frequency Range 1}
\newacronym{fr2}{FR2}{Frequency Range 2}
\newacronym{fr3}{FR3}{Frequency Range 3}
\newacronym{fss}{FSS}{Fixed Satellite Service}
\newacronym{geo}{GEO}{geostationary Earth orbit}
\newacronym{ggs}{GGS}{Guarded Greedy Scheduler}
\newacronym{gnb}{gNB}{next-generation NodeB}
\newacronym{gpu}{GPU}{graphics processing unit}
\newacronym{harq}{HARQ}{Hybrid Automatic Repeat Request}
\newacronym{ibcp}{IB-CP}{Interference-Budget Convex Problem}
\newacronym{ibp}{IBP}{Interference-Budget Problem}
\newacronym{if}{IF}{intermediate frequency}
\newacronym{in}{$I/N$}{interference-to-noise ratio}
\newacronym{itur}{ITU-R}{International Telecommunication Union Radiocommunication Sector}
\newacronym{kb}{KB}{knowledge base}
\newacronym{kkt}{KKT}{Karush-Kuhn-Tucker}
\newacronym{kpm}{KPM}{Key Performance Measurement}
\newacronym{llm}{LLM}{Large Language Model}
\newacronym{lna}{LNA}{low-noise amplifier}
\newacronym{lo}{LO}{local oscillator}
\newacronym{lp}{LP}{linear program}
\newacronym{mac}{MAC}{Medium Access Control}
\newacronym{mcs}{MCS}{Modulation and Coding Scheme}
\newacronym{mdp}{MDP}{Markov Decision Process}
\newacronym{ml}{ML}{machine learning}
\newacronym{nic}{NIC}{network interface card}
\newacronym{nr}{NR}{New Radio}
\newacronym{nrdz}{NRDZ}{National Radio Dynamic Zone}
\newacronym{nrt}{near-RT}{near-real-time}
\newacronym{oai}{OAI}{OpenAirInterface}
\newacronym{oam}{OAM}{Operations, Administration, and Maintenance}
\newacronym{ocu}{O-CU}{O-RAN Central Unit}
\newacronym{odu}{O-DU}{O-RAN Distributed Unit}
\newacronym{ofdm}{OFDM}{Orthogonal Frequency-Division Multiplexing}
\newacronym{oran}{O-RAN}{Open Radio Access Network}
\newacronym{oru}{O-RU}{O-RAN Radio Unit}
\newacronym{ota}{OTA}{over-the-air}
\newacronym{pa}{PA}{power amplifier}
\newacronym{pdcp}{PDCP}{Packet Data Convergence Protocol}
\newacronym{pdu}{PDU}{Packet Data Unit}
\newacronym{phy}{PHY}{physical layer}
\newacronym{plmn}{PLMN}{Public Land Mobile Network}
\newacronym{prach}{PRACH}{Physical Random Access Channel}
\newacronym{prb}{PRB}{Physical Resource Block}
\newacronym{psd}{PSD}{power spectral density}
\newacronym{ptp}{PTP}{precision time protocol}
\newacronym{pusch}{PUSCH}{Physical Uplink Shared Channel}
\newacronym{qam}{QAM}{Quadrature Amplitude Modulation}
\newacronym{qpsk}{QPSK}{Quadrature Phase Shift Keying}
\newacronym{ra}{RA}{random access}
\newacronym{ran}{RAN}{Radio Access Network}
\newacronym{rcpo}{RCPO}{Reward-Constrained Policy Optimization}
\newacronym{rdz}{RDZ}{Radio Dynamic Zone}
\newacronym{rf}{RF}{radio frequency}
\newacronym{ric}{RIC}{RAN Intelligent Controller}
\newacronym{rl}{RL}{reinforcement learning}
\newacronym{rrc}{RRC}{Radio Resource Control}
\newacronym{rsrp}{RSRP}{Reference Signal Received Power}
\newacronym{rtt}{RTT}{round-trip time}
\newacronym{sac}{SAC}{Soft Actor-Critic}
\newacronym{scs}{SCS}{subcarrier spacing}
\newacronym{sctp}{SCTP}{Stream Control Transmission Protocol}
\newacronym{sdr}{SDR}{software-defined radio}
\newacronym{sgs}{SGS}{Sensing-driven Greedy Scheduler}
\newacronym{sinr}{SINR}{signal-to-interference-plus-noise ratio}
\newacronym{siso}{SISO}{Single-Input Single-Output}
\newacronym{smf}{SMF}{Session Management Function}
\newacronym{smo}{SMO}{Service Management and Orchestration}
\newacronym{snr}{SNR}{signal-to-noise ratio}
\newacronym{sriov}{SR-IOV}{single-root I/O virtualization}
\newacronym{ssb}{SSB}{Synchronization Signal Block}
\newacronym{tcp}{TCP}{Transmission Control Protocol}
\newacronym{td}{TD}{temporal difference}
\newacronym{tdd}{TDD}{Time Division Duplex}
\newacronym{tpc}{TPC}{Transmit Power Control}
\newacronym{udp}{UDP}{User Datagram Protocol}
\newacronym{ue}{UE}{user equipment}
\newacronym{ul}{UL}{Uplink}
\newacronym{ulsch}{UL-SCH}{Uplink Shared Channel}
\newacronym{upf}{UPF}{User Plane Function}
\newacronym{urllc}{URLLC}{Ultra-Reliable Low-Latency Communications}
\newacronym{vf}{VF}{virtual function}
\newacronym{wrc23}{WRC-23}{World Radiocommunication Conference 2023}

\newcommand{\genesis}{\textsc{Genesis}}
\newcommand{\ardz}{A-RDZ}

\usepackage{enumitem}

\usepackage{fancyhdr}

\fancypagestyle{firstpage}{
	\fancyhf{}
	
	\fancyhead[C]{%
		\fbox{\parbox{\dimexpr\textwidth-2\fboxsep-2\fboxrule\relax}{
                \centering
				\footnotesize\normalfont
				This paper has been accepted for publication at IEEE MILCOM Workshops 2026. This is the author’s accepted version of the paper.
		}}
	}
}

\begin{document}
	\bstctlcite{IEEEexample:BSTcontrol}
    
	\title{Agentic RDZ: Autonomous Zone Management with AI Agents and an FR3 Coexistence Use Case}
	
	\author{
		\IEEEauthorblockN{Minh Dat Nguyen, Gabriele Gemmi, Tamerlan Aghayev, Paolo Testolina, Michele Polese, Tommaso Melodia}
		\IEEEauthorblockA{Institute for Intelligent Networked Systems, Northeastern University, Boston, MA, U.S.A.
			\\\{minhd.nguyen, g.gemmi, aghayev.t, p.testolina, m.polese, melodia\}@northeastern.edu
            \vspace{-0.5cm}}
		\thanks{This work was supported by the U.S. National Science Foundation under Grants CNS-2434081 and OSI-2431961.}
		
	}
	
	\maketitle
    \thispagestyle{firstpage}
	
	\begin{abstract}
		\Glspl{rdz} allow wireless experiments to operate outside conventional spectrum regulations while continuously guaranteeing protection for incumbent users. Existing \gls{rdz} prototypes automate this task procedurally, through handcrafted rules and predefined workflows, and become brittle when experiments encounter hardware impairments, user workflows and devices, or interference mechanisms not anticipated at design time.
		This paper introduces the \emph{agentic \gls{rdz}} (\ardz{}), which, to the best of our knowledge, is the first \gls{rdz} realization in which agents use \glspl{llm} to perform spectrum management, experiment management, policy interpretation, and zone orchestration. Built on the \genesis{} agentic framework, the architecture pairs autonomous reasoning with a deterministic policy gate and \gls{nrt} reflexes, so that agents can improve outcomes but never weaken the zone's protection guarantee.
		We validate the \ardz{} on a hardware-in-the-loop \gls{fr3} (7.125--24.25\,GHz) \gls{oran} testbed in which a 5G \gls{nr} experiment coexists with an emulated \gls{fss} earth-station incumbent. In an end-to-end use case, the monitoring agent detects an emission violation from live spectrum evidence, 
		the orchestrator selects a mitigation that restores the interference budget while keeping the experiment running, and the action is applied and verified through the \gls{oran} control plane.
		We report the detection-to-mitigation latency decomposition and discuss the practical limits of agentic operation, including non-deterministic reasoning and decision-to-action translation.
	\end{abstract}
	
	
	\glsresetall
	
	\section{Introduction}
	\label{sec:intro}
	
	Future 6G systems are expected to exploit the \gls{fr3} upper mid-band (7.125--24.25\,GHz) for wider bandwidths than conventional sub-7\,GHz systems~\cite{NYU_FR3_vision,FR3:6G25}. However, \gls{fr3} is already occupied by incumbents such as \gls{fss} earth stations and federal systems, making coexistence a key deployment challenge. 
	\textcolor{black}{Characterizing that coexistence requires \gls{ota} measurements with real hardware and protocol stacks to account for non-ideal \gls{rf} behaviors and propagation. 
		Experimental authorization is granted against pre-declared parameters at a fixed site.
        Highly-dynamic or wide-ranging tests, however, have requirements that are hard to predict, or change based on the outcome of the experiment itself.
        In that case, obtaining a new license might significantly slow down the experiment pace and the scientific output.}
	
	\Glspl{rdz} address this limitation by providing controlled environments for spectrum experimentation, enabling wireless systems to operate under dedicated experimental authorizations while continuously protecting incumbent users~\cite{Zheleva2023RDZ}.
	Existing systems, including OpenZMS on POWDER~\cite{OpenZMS2025,PowderRDZ2024}, COSMOS~\cite{CosmosZMS2026}, and FlexRDZ~\cite{FlexRDZ2023}, automate spectrum, experiment, and policy management through predefined rules, workflows, or task models. 
	\textcolor{black}{Their coverage is therefore bounded by what the designer explicitly specified: a hardware impairment or interference mechanism outside that enumeration has no matching rule, and the zone's remaining response is to terminate the experiment.}
	
	Recent agentic \gls{ai} systems provide complementary capabilities: they can reason over heterogeneous evidence and invoke tools~\cite{yao2022react}, recover from failures through self-correction~\cite{noah2023reflexion}, and coordinate multi-step operations~\cite{wu2024autogen}. \genesis{}~\cite{GENESIS}, for example, demonstrated multi-agent autonomous operation of \gls{oran} testbeds. These capabilities motivate applying agentic \gls{ai} to \gls{rdz} operation, while retaining deterministic mechanisms for safety-critical protection.
	
	Motivated by these observations, we propose the \emph{agentic \gls{rdz}} (\ardz{}), which, to the best of our knowledge, is the first extension of the conventional \gls{rdz} architecture with coordinated \gls{ai} agents for spectrum, experiment, and policy management and zone orchestration.
    The architecture supports AI-based decisions, guard-railed by deterministic policies and real-time protection mechanisms.
    We validate it on a hardware-in-the-loop \gls{fr3} \gls{oran} testbed, where a 5G \gls{nr} experiment coexists with an emulated \gls{fss} incumbent.
	\textcolor{black}{During the experiment, a controlled configuration change pushes the unwanted emissions into the protected sub-band.
    The monitoring agent detects that the measured interference exceeds the protection limit and identifies the inverting frequency plan as the cause, rather than the transmitter's in-band configuration.
    The orchestrator then selects a \gls{prb}-region mask that brings the interference back within the limit, while the experiment continues to run.}
	The main contributions are:
	\begin{itemize}
		\item We introduce the concept of an \emph{agentic \gls{rdz}} (\ardz{}), extending conventional RDZ functions with coordinated AI agents while retaining deterministic safety mechanisms.
		
		\item We integrate \ardz{} with the \genesis{} framework and a hardware-in-the-loop \gls{fr3} \gls{oran} coexistence testbed, demonstrating agentic detection, mitigation selection, and policy-gated execution.
		
		\item We evaluate the end-to-end enforcement loop on several metrics, including latency, protection compliance, experiment utility, and the role of deterministic safeguards under agentic operation.
	\end{itemize}
	
	\section{RDZ Operational Challenges and Agentic Support}
	\label{sec:challenges}
	
	An \gls{rdz} consists of three functional subsystems coordinated by a decision engine~\cite{Zheleva2023RDZ}: 
	Spectrum Management, Experiment Management, and Policy Management. We summarize the key operational challenges of each subsystem and discuss how an agentic realization complements conventional procedural automation. 
	Table~\ref{tab:mapping} summarizes the mapping.
	
	\begin{table}[t]
		\caption{RDZ operational challenges mapped to agentic capabilities and to the deterministic guards that bound agent behavior.}
		\label{tab:mapping}
		\centering
		\footnotesize
		\begin{tabular}{@{}p{1.7cm}p{2.9cm}p{3.3cm}@{}}
			\toprule
			\textbf{RDZ function} &
			\textbf{Deterministic safeguard} 
			&
			\textbf{Agentic capability} \\
			\midrule
			Spectrum management &
            Reflex threshold test and pre-configured protective action &
			Evidence-based anomaly reasoning and predictive violation detection \\
			\addlinespace
			
			Experiment management &
			Grant validation before execution &
			Autonomous deployment, diagnosis, and recovery \\
			\addlinespace
			
			Policy management &
			Non-overridable protection limits &
			Reasoning over structured and natural-language policies \\
			\addlinespace
			
			Decision engine &
			\gls{nrt} reflexes (10--1000 ms); agents (1--200 s) &
			Cross-subsystem coordination and mitigation planning \\
			\bottomrule
		\end{tabular}
        \vspace{-10pt}
	\end{table}
	
	
	\begin{itemize}[leftmargin=1em, labelwidth=0.6em, itemindent=0pt, labelsep=0.3em, align=left]
		
		\item \textbf{Spectrum Management:}
		Spectrum management must detect interference violations and anticipate harm to protected incumbents. Conventional systems rely on threshold checks derived from propagation and emission models, which may fail when experimental hardware behaves unexpectedly.
		\textcolor{black}{
			An agent can instead correlate spectrum measurements with experiment configurations and historical observations to attribute anomalous emissions and predict violations before protection limits are exceeded. 
            It can also assess whether a measured exceedance causes harmful interference at the incumbent. Each individual computation is scriptable and is already implemented in existing systems. The agent is useful when the script has no matching case: it can assess an incumbent scenario, geometry, or interference path not covered by the encoded rule set, and defers to the deterministic threshold when no such assessment can be made.}
		
		\item \textbf{Experiment Management:}
		\label{sec:chal-experiment}
		Experiments execute on shared radio and compute infrastructure, requiring reliable deployment, monitoring, recovery, and cleanup. While scripts automate nominal workflows, diagnosing failed deployments often requires reasoning over logs, system state, and configurations. 
		\textcolor{black}{
			Moreover, agentic management supports experiment workflows outside the encoded set, including mapping a new workflow onto the spectrum-management requirements it implies. Deterministic policy checks keep all lifecycle actions within the experiment's authorized spectrum grant.}
		
		\item \textbf{Policy Management:}
		\label{sec:chal-policy}
		Zone policies combine structured spectrum rules with natural-language regulatory documents and coordination agreements. Procedural systems require these policies to be manually encoded into machine-parsable text or code before enforcement. 
		A \gls{llm}-based agent can instead interpret heterogeneous policy documents, directly evaluate natural-language ones, evaluate experiment requests, and recommend the least-disruptive compliant mitigation. Hard protection constraints remain enforced by a deterministic policy gate.
		
		\item \textbf{Decision Engine:}
		\label{sec:chal-decision}
		The decision engine coordinates the three subsystems while satisfying stringent protection timescales. 
		Since \gls{llm} inference cannot guarantee sub-second response, the \ardz{} adopts a two-timescale architecture: deterministic \gls{nrt} reflexes provide immediate protection, while agents operate at slower timescales to diagnose anomalies, select mitigations, and coordinate system-wide actions.
		
		
	\end{itemize}
    Agents can thus increase the autonomy and automation of \glspl{rdz} by interpreting heterogeneous information and recovering from unforeseen operating conditions.
    However, agentic operation introduces two key challenges: non-deterministic reasoning and imperfect decision-to-action translation. 
    Accordingly, safety guarantees are required.
    Deterministic policy gates and \gls{nrt} reflexes, enforced through typed interfaces and action verification routines, limit the agent errors.
	
	\section{Agentic RDZ Architecture}
	\label{sec:architecture}
	
	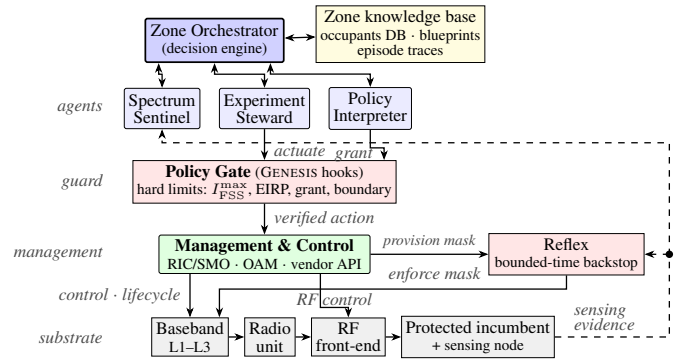
\begin{figure}[t]
		\centering
		\resizebox{\linewidth}{!}{
\begingroup
\footnotesize
\begin{tikzpicture}[
	font=\footnotesize,
	node distance=3mm and 3mm,
	agent/.style={draw, rounded corners=1pt, fill=blue!8, align=center,
		minimum height=6.5mm, inner sep=2pt},
	orch/.style={draw, rounded corners=1pt, fill=blue!18, align=center,
		minimum height=7mm, inner sep=2.5pt, thick},
	kb/.style={draw, fill=yellow!15, align=center, minimum height=6.5mm,
		inner sep=2pt},
	det/.style={draw, fill=red!10, align=center, minimum height=6mm,
		inner sep=2pt},
	mgmt/.style={draw, rounded corners=1pt, fill=green!12, align=center,
		minimum height=6mm, inner sep=2pt},
	hw/.style={draw, fill=gray!12, align=center, minimum height=6mm,
		inner sep=2pt},
	lbl/.style={font=\footnotesize\itshape, gray!60!black},
	flow/.style={-{Stealth[length=1.6mm]}, semithick},
	evid/.style={-{Stealth[length=1.6mm]}, semithick, dashed},
	bidir/.style={{Stealth[length=1.6mm]}-{Stealth[length=1.6mm]}, semithick}
	]
	\node[orch] (orch) {Zone Orchestrator\\[-1pt]\scriptsize(decision engine)};
	\node[agent, below left=4mm and -9mm of orch] (sent)
	{Spectrum\\[-1pt]Sentinel};
	\node[agent, right=3mm of sent] (stew)
	{Experiment\\[-1pt]Steward};
	\node[agent, right=3mm of stew] (poli)
	{Policy\\[-1pt]Interpreter};
	\node[kb, right=5mm of orch, yshift=1mm] (kb)
	{Zone knowledge base\\[-1pt]\scriptsize occupants DB $\cdot$ blueprints\\[-2pt]\scriptsize episode traces};
	\draw[bidir] (sent.north) -- ++(0,2.2mm) -| ($(orch.south)+(-9mm,0)$);
	\draw[bidir] (stew.north) -- ++(0,2.2mm) -| (orch.south);
	\draw[bidir] (poli.north) -- ++(0,2.1mm) -| ($(orch.south)+(9mm,0)$);
	\draw[bidir] (orch) -- (kb);
	\node[det, below=5mm of stew, minimum width=34mm] (gate)
	{\textbf{Policy Gate} {\scriptsize(\textsc{Genesis} hooks)}\\[-1pt]\scriptsize hard limits: $I^{\max}_{\mathrm{FSS}}$, EIRP, grant, boundary};
	\node[mgmt, below=5mm of gate, minimum width=34mm] (mctrl)
	{\textbf{Management\ \& Control}\\[-1pt]\scriptsize RIC/SMO $\cdot$ OAM $\cdot$ vendor API};
	\node[det, right=19mm of mctrl, minimum width=25mm] (reflex)
	{Reflex\\[-1pt]\scriptsize bounded-time backstop};
	\draw[flow] (stew) -- (gate)
	node[lbl, pos=0.7, right, xshift=0.5mm]{actuate};
	\draw[flow] (poli.south) -- ++(0,-2.9mm) -|
	node[lbl, pos=0.3, below=-1mm, xshift=-4mm]{grant} ($(gate.north east)+(-2mm,0)$);
	\draw[flow] (gate) -- node[lbl, right, xshift=0.5mm]{verified action} (mctrl);
	\draw[flow] (mctrl) --
	node[lbl, above, font=\scriptsize\itshape]{provision mask} (reflex);
	\node[hw, below=7mm of mctrl, xshift=-12mm] (gnb)
	{Baseband\\[-1pt]\scriptsize L1--L3};
	\node[hw, right=2.5mm of gnb] (oru) {Radio\\[-1pt]unit};
	\node[hw, right=2.5mm of oru] (pirad)
	{RF\\[-1pt]front-end};
	\node[hw, right=2.5mm of pirad] (fss)
	{Protected incumbent\\[-1pt]\scriptsize + sensing node};
	\draw[flow] (gnb) -- (oru);
	\draw[flow] (oru) -- (pirad);
	\draw[flow] (pirad) -- (fss);
	
	\draw[flow] (mctrl.south -| gnb.north) --
	node[lbl, pos=0.5, left, xshift=-0.5mm]{control $\cdot$ lifecycle} (gnb.north);
	\draw[flow] ($(mctrl.south)+(9mm,0)$) -- ++(0,-5.4mm) -|
	node[lbl, pos=0.25, above=-0.6mm]{RF control} (pirad.north);
	\coordinate (t0) at ($(reflex.south)+(0,-2.2mm)$);
	\coordinate (t2) at ($(gnb.north east)+(-1.5mm,0)$);
	\draw[flow] (reflex.south) -- (t0) --
	node[lbl, pos=0.38, above]{enforce mask} (t0 -| t2) -- (t2);
	\coordinate (rm) at ($(reflex.east)+(4mm,0)$);
	\draw[evid] (fss.east) -- (rm |- fss.east) |-
	($(sent.south)+(0,-2.2mm)$) -| (sent.south);
	\node[lbl, align=left, anchor=west, xshift=8mm, above=0.1mm] at (fss.east) {sensing\\[-2pt]evidence};
	\coordinate (rj) at (rm |- reflex.east);
	\fill (rj) circle (0.6mm);
	\draw[evid] (rj) -- (reflex.east);
	\coordinate (lm) at ($(mctrl.west)+(-8mm,0)$);
	\node[lbl, anchor=east] at (lm |- sent.west) {agents};
	\node[lbl, anchor=east] at (lm |- gate.west) {guard};
	\node[lbl, anchor=east] at (lm |- mctrl.west) {management};
	\node[lbl, anchor=east] at (lm |- gnb.west) {substrate};
\end{tikzpicture}
\endgroup}
		\caption{The agentic \gls{rdz} architecture: \genesis{} agents (top) realize the \gls{rdz} functional components of~\cite{Zheleva2023RDZ}.}
		\label{fig:arch}
        \vspace{-3mm}
	\end{figure}

	Figure~\ref{fig:arch} shows the \ardz{} architecture. 
	\textcolor{black}{We implement it by building on \genesis{}~\cite{GENESIS}, an agentic framework for \gls{oran} testbeds that provides domain-scoped agents, reusable skills, and deterministic \emph{hooks} that can block unsafe actions before network execution, to implement and extend the \gls{rdz} functional components of~\cite{Zheleva2023RDZ}}
	
	The architecture is organized into four planes. An \emph{agent plane} performs the reasoning functions of the zone; a \emph{guard plane} enforces the zone's guarantees through deterministic protections; a \emph{management plane} translates verified decisions into network configuration; and a \emph{substrate plane} includes the managed radio system, the protected incumbent, and the sensing nodes that observe it. 
	\textcolor{black}{The management plane provides two control paths: an agentic path that applies only decisions verified by the guard plane, and a deterministic \emph{reflex} that applies a pre-configured protective action when the protection threshold is crossed, independently of agent liveness.} 
	

    \textcolor{black}{The management plane is the sole component permitted to write to the substrate, making the architecture independent of the 
    interface used to reach it: \gls{oran} E2/O1, 3GPP \gls{oam}, or a device-specific \gls{api}.
	The zone knowledge base stores the occupants database, the measurement and policy blueprints, and the enforcement history.
    It may be specific to a zone, or federated with an experimenter's knowledge base.}
	

	\begin{itemize}[leftmargin=1em, labelwidth=0.6em, itemindent=0pt, labelsep=0.3em, align=left]
		\item \textbf{Specialist Agents:}
		\label{sec:arch-agents}
		\textcolor{black}{Following the \genesis{} design, in which each agent is scoped to a single domain and given only the tools and authority required for that domain, three specialist agents realize the core \gls{rdz} subsystems.}
		\begin{itemize}
			\item \textit{Spectrum Sentinel (spectrum management):}
			\textcolor{black}{
				The Spectrum Sentinel manages the sensing nodes and the measurement blueprint.
                It implements a traditional interference threshold test 
				as a \emph{reflex}: a deterministic check that requires no reasoning. 
                The Sentinel agent extends it by testing whether the observed spectrum matches the declared configuration. From the carrier, occupied bandwidth, \gls{prb} allocation, and frequency plan, it determines where emitted power should appear. A change in the occupied region indicates a configuration change; energy in the adjacent sub-band predicted by the injection side indicates a frequency-plan artifact; and energy where no resources are allocated indicates that the transmitter is not operating as declared.
				It also assesses whether a limit breach can cause actual harm at the incumbent, and flags emissions inconsistent with the declared configuration while still within budget.
                Its outputs are detected and predicted violations.}
			
			\item \textit{Experiment Steward (experiment management):}
			\textcolor{black}{
				The Experiment Steward manages the occupants database and experiment lifecycle. 
                It instantiates an experiment from its spectrum grant and declared configuration, then verifies the resulting emissions against the grant.
                During the run, it applies reconfigurations, recovers from common failures, and executes orchestrator-selected mitigation actions. It tears the experiment down at completion or following an enforcement decision.
				All lifecycle operations are validated by the policy gate before execution.}
			
			\item \textit{Policy Interpreter (policy management):}
			The Policy Interpreter manages the zone policy blueprint, including protection criteria, coordination agreements, and admission rules. It translates experiment requests into enforceable spectrum grants, evaluates mitigation actions for policy compliance, and maintains an auditable record of policy decisions.
			
		\end{itemize}

		\item \textbf{Zone Orchestrator (decision engine):}
		\label{sec:arch-orch}
        \textcolor{black}{
        The Zone Orchestrator implements the \gls{rdz} decision engine. It is a deterministic component rather than an agent: it has no \gls{llm} of its own and invokes the specialist agents as typed functions. 
        On a violation, it collects the attribution, candidate mitigations with projected post-action interference and disruption cost, and the protection target, then selects the lowest-cost candidate that meets the target. The selected action is delegated through the policy gate and verified by an independent measurement. 
        The agents provide attribution and candidate mitigations; the orchestrator makes the final selection. At admission, it also derives the protection threshold and pre-computed reflex action from the grant. All decisions and outcomes are recorded in the knowledge base.
        }

		\item \textbf{Guard Plane:}
		\label{sec:arch-gate}
		All actuation originating in the agent plane passes through a \emph{policy gate}: a deterministic validator holding the zone's hard limits, the incumbent interference budget $I^{\max}_{\mathrm{FSS}}$ derived from the applicable protection criterion, absolute \gls{eirp} caps, the frequency grant of each experiment, and the geographic boundary.
		\textcolor{black}{
			The gate rejects any action that would result in a violation, and only actions it has verified are admitted to the management plane.
            It follows the reference-monitor model~\cite{Anderson1972} and, for learned policies, the shielding pattern~\cite{Alshiekh2018}, in which a correct-by-construction monitor overrides any agent action violating a safety specification.
            In the \genesis{} substrate, this is realized as hooks that fire out-of-band of the agents' reasoning loop and can hard-block an action before it reaches the network~\cite{GENESIS}.
            The gate deterministically evaluates the expected outcome from the action parameters and the current calibrated measurement: a \gls{prb} mask removes the measured contribution of the selected \glspl{prb}, while a power offset scales the emission by a fixed factor. The agent only proposes the action; the gate independently computes and checks its expected outcome.}
		
		\item \textbf{Management Plane:}
		\label{sec:arch-mgmt}
		The management plane converts a verified action into concrete configuration on the substrate and arbitrates between the two timescales at which the zone acts. 
		\textcolor{black}{
			A verified \gls{prb}-mask action, for example, becomes an \gls{oran} \gls{ric} control message in our testbed, a scheduler configuration change under 3GPP \gls{oam}, or a front-end gain reduction where no network-control path exists.}
		It is the sole point at which the architecture writes to the network and therefore the only place specialized to a particular technology.
		\textcolor{black}{
			The management plane also hosts the \emph{reflex}: a pre-configured, bounded-time protection path that evaluates the monitored interference budget against the protection threshold established at experiment admission and applies a pre-computed protective action, such as a \gls{prb} mask, when the monitored budget is exceeded. It operates independently of agent liveness and performs no \gls{llm} inference, so its response time is bounded by the sensing cadence and the control interval of the underlying network rather than by reasoning latency.}


    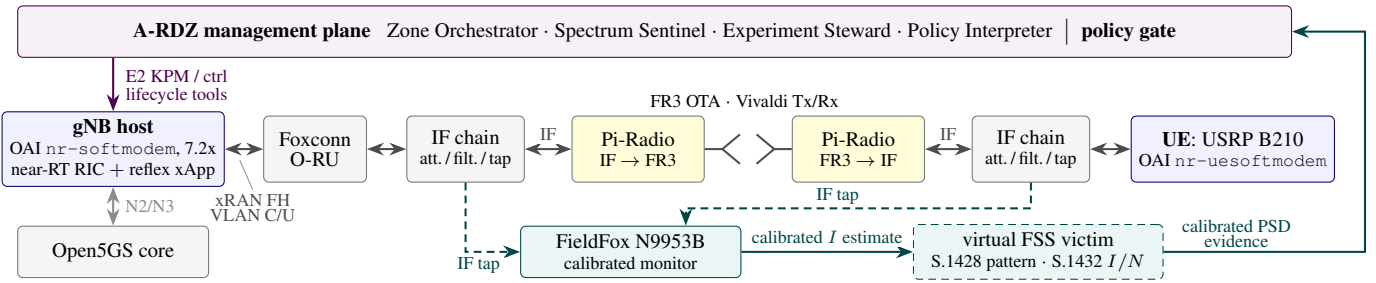
\begin{figure*}[t]
        \centering
		\resizebox{\textwidth}{!}{

\begin{tikzpicture}[
	font=\footnotesize,
	>=Stealth,
	base/.style   = {draw, rounded corners=2pt, align=center, inner sep=3pt,
		minimum height=9mm, fill=white, line width=0.4pt},
	host/.style   = {base, fill=blue!6,   draw=blue!45!black},
	rfblk/.style  = {base, fill=black!4,  draw=black!55},
	piradio/.style= {base, fill=yellow!18, draw=black!55},
	meas/.style   = {base, fill=teal!8,   draw=teal!55!black, minimum height=7mm},
	virt/.style   = {meas, densely dashed},
	plane/.style  = {base, fill=violet!6, draw=violet!55!black, minimum height=7mm},
	note/.style   = {font=\scriptsize, align=center, inner sep=1.5pt},
	slbl/.style   = {font=\scriptsize, fill=white, inner sep=1.5pt},
	sig/.style    = {<->, thick, black!70},
	tap/.style    = {->, densely dashed, teal!55!black, thick},
	ev/.style     = {->, teal!55!black, thick},
	ctrl/.style   = {->, violet!60!black, thick},
	]
	
	\node[plane, minimum width=17.35cm] (mgmt) at (8.775,1.6)
	{\textbf{A-RDZ management plane}\;\;
		Zone Orchestrator $\cdot$ Spectrum Sentinel $\cdot$ Experiment Steward $\cdot$
		Policy Interpreter \;$\big|$\; \textbf{policy gate}};
	
	\node[host,  minimum width=2.6cm] (gnb)    at (1.40,0)
	{\textbf{gNB host}\\[-1pt]{\scriptsize OAI \texttt{nr-softmodem}, 7.2x}\\[-1pt]
		{\scriptsize near-RT RIC $+$ reflex xApp}};
	
	\node[rfblk, minimum width=1.4cm] (oru)    at (4.15,0)  {Foxconn\\[-1pt]O-RU};
	
	\node[rfblk, minimum width=1.6cm] (chainA) at (6.20,0)
	{IF chain\\[-1pt]{\scriptsize att.\,/\,filt.\,/\,tap}};
	
	\node[piradio, minimum width=1.8cm] (pr1)    at (8.55,0)
	{Pi-Radio\\[-1pt]{\scriptsize IF $\to$ FR3}};
	
	\node[piradio, minimum width=1.8cm] (pr2)    at (11.55,0)
	{Pi-Radio\\[-1pt]{\scriptsize FR3 $\to$ IF}};
	
	\node[rfblk, minimum width=1.6cm] (chainB) at (13.90,0)
	{IF chain\\[-1pt]{\scriptsize att.\,/\,filt.\,/\,tap}};
	
	\node[host,  minimum width=2.4cm] (ue)     at (16.65,0)
	{\textbf{UE}: USRP B210\\[-1pt]{\scriptsize OAI \texttt{nr-uesoftmodem}}};
	
	\node[host, fill=black!4, draw=black!55, minimum width=2.6cm, minimum height=7mm]
	(core) at (1.40,-1.4) {Open5GS core};
	
	\node[meas, minimum width=3.0cm] (ff) at (8.45,-1.4)
	{FieldFox N9953B\\[-1pt]{\scriptsize calibrated monitor}};
	
	\node[virt, minimum width=3.4cm] (victim) at (14.00,-1.4)
	{virtual FSS victim\\[-1pt]{\scriptsize S.1428 pattern $\cdot$ S.1432 $I/N$}};
	
	\draw[sig] (gnb) -- (oru);
	\node[note, black!70] (fhlbl) at (3.30,-0.80) {xRAN FH\\[-2pt]VLAN C/U};
	\draw[black!45, line width=0.3pt] (fhlbl.north) -- (3.125,-0.15);
	\draw[sig] (oru) -- (chainA);
	\draw[sig] (chainA) -- node[slbl, above=1.5pt]{IF} (pr1);
	\draw[sig] (pr2) -- node[slbl, above=1.5pt]{IF} (chainB);
	\draw[sig] (chainB) -- (ue);
	\draw[<->, thick, black!45] (gnb) -- node[slbl, right=1mm]{N2/N3} (core);
	
	\coordinate (a1) at ($(pr1.east)+(0.24,0)$);
	\coordinate (a2) at ($(pr2.west)-(0.24,0)$);
	\draw[black!70, thick] (pr1.east) -- (a1)
	($(a1)+(0.24,0.20)$) -- (a1) -- ($(a1)+(0.24,-0.20)$);
	\draw[black!70, thick] (pr2.west) -- (a2)
	($(a2)-(0.24,-0.20)$) -- (a2) -- ($(a2)-(0.24,0.20)$);
	\node[note] at (10.0,0.65) {FR3 OTA $\cdot$ Vivaldi Tx/Rx};
	
	\draw[tap] (chainA.south) |- node[slbl, pos=0.62, below]{IF tap} (ff.west);
	\draw[tap] (chainB.south) -- ++(0,-0.35) -| node[slbl, pos=0.28, above]{IF tap}
	($(ff.north)+(0.75,0)$);
	\draw[ev]  (ff.east) -- node[slbl, above=1.5pt]{calibrated $I$ estimate} (victim.west);
	
	\draw[ctrl] (mgmt.south -| gnb) -- node[slbl, right=3pt, align=left, pos=0.55]
	{E2 KPM / ctrl\\[-1pt]lifecycle tools} (gnb.north);
	\draw[ev] (victim.east) -- (18.450,-1.4) -- (18.450,1.6) -- (mgmt.east);
	\node[note, teal!55!black, anchor=south] at (16.70,-1.4)
	{calibrated PSD\\[-2pt]evidence};
	
\end{tikzpicture}}
		\caption{\acrshort{fr3} \acrshort{oran} testbed and its exposure to the \ardz{} management plane.}
		\label{fig:testbed}
		\vspace{-4mm}
	\end{figure*}
		
		\item \textbf{Substrate Plane and Sensing:}
		\label{sec:arch-substrate}
		The substrate plane comprises the managed radio, processing stack, RF front-end, protected incumbent, and sensing nodes. The management plane exposes \emph{network-control} actions (e.g., \gls{prb} masking, power adjustment, and \gls{mcs} capping), \emph{lifecycle-control} actions (container, process, and configuration management), and \emph{RF-control} actions such as transmit-gain adjustment.
		Sensing is a northbound observation path: calibrated \gls{psd} captures reach the Spectrum Sentinel without traversing the management plane. Zone-boundary sensing is left to future work in this single-link realization.
		
	\end{itemize}
	
	
	
	\section{Experimental Setup and \genesis{}/FR3 Integration}
	\label{sec:setup}
	
	\subsection{FR3 O-RAN Testbed}
	\label{sec:setup-testbed}

	The experiment managed by the \ardz ~(Fig.~\ref{fig:testbed}) is a standalone 5G \gls{nr} network built on \gls{oai}~\cite{OAI}. 
	The \gls{gnb} runs \texttt{nr-softmodem} with a 7.2x functional split in a Docker container, driving a Foxconn \gls{oru} over xRAN fronthaul via an Intel X710 \gls{nic}.
	The cell operates in band n78 with 106 \glspl{prb} at 30\,kHz \gls{scs} ($\mu\!=\!1$) and $f_{\mathrm{IF}}\!=\!3.58$\,GHz.
    The \gls{fr1}/\gls{if} path uses calibrated attenuators, filters, and splitters with monitoring taps for repeatable coupling.
	A pair of Pi-Radio up/down-conversion boards~\cite{PiRadio_6G_Proto} translates the \gls{if} to \gls{fr3} using a common $f_{\mathrm{LO}}\!=\!13.58$\,GHz with high-side injection, inverting the spectrum and placing the dominant emission skirt in the lower-adjacent sub-band.
	The \gls{fr3} segment forms a short indoor \gls{ota} link between directive Vivaldi antennas, with the fixed geometry included in the end-to-end calibration. A USRP B210 running \texttt{nr-uesoftmodem} serves as the \gls{ue} behind the down-converter, while Open5GS provides the core network on a separate host.

	\subsection{Incumbent Emulation and Sensing}
	
	The managed incumbent is an \emph{emulated} \gls{fss} earth-station receiver. Because the USRP B210 cannot operate at \gls{fr3}, a FieldFox N9953B serves as the physical sensing device, while the victim-receiver characteristics are applied analytically to the calibrated captures. 
	The measured victim-band power is mapped from the \gls{if} through the verified frequency plan, referred to the victim plane using the calibrated Vivaldi antenna and RF chain, and weighted according to the \gls{itur} S.1428 off-axis gain~\cite{ITU_S1428}. 
	With $N=kT_{\mathrm{sys}}B$, where $k$ is the Boltzmann constant, $T_{\mathrm{sys}}$ is the system noise temperature, and $B$ is the reference bandwidth, the \gls{itur} S.1432 criterion $I/N\leq-10$\,dB~\cite{ITU_S1432} defines the interference budget $I_{\mathrm{FSS}}^{\max}$ enforced by the policy gate. 
    \textcolor{black}{
    FieldFox captures are timestamped and delivered to the Spectrum Sentinel; their capture cadence contributes to the enforcement-loop latency.
    }
	
	\vspace{-1mm}
	\subsection{Agent Integration}
	
	The \ardz{} management plane runs off-testbed and reaches the infrastructure through three tool families, each exposed to the agents as a typed, schema-validated function and protected by the policy gate:
	\begin{itemize}
		\item \emph{\gls{ric} tools}: \gls{kpm} subscription for per-\gls{ue}/per-slice telemetry and \gls{ran}-control actions (\gls{prb}-region mask, power offset, and \gls{mcs} cap) issued via an xApp on the \gls{nrt} \gls{ric} using FlexRIC~\cite{FlexRIC}.
		
		\item \emph{Lifecycle tools}: container control (deploy from the committed \gls{gnb} image, stop, commit, log retrieval), configuration templating for the \gls{oai} and \gls{oru} config files, and host hygiene actions (fronthaul \gls{vf} re-initialization with the pinned MAC address, stale hugepage-map cleanup) that the Steward runs before any deployment.
		
		\item \emph{Sensing tools}: triggered and periodic capture from the FieldFox, which also realizes the virtual victim observation point, returning calibrated \glspl{psd} and scalar budget estimates.
	\end{itemize}
	
	The reflex is an xApp configured at experiment admission: it evaluates the interference-budget estimate at each control interval ($\Delta t \!=\! 100$\,ms) and applies a pre-computed protective \gls{prb} mask if the instantaneous estimate exceeds $I_{\mathrm{FSS}}^{\max}$, independently of agent liveness.
	
	\vspace{-1mm}
	\subsection{Use Case: Detection-to-Mitigation}
	\label{sec:use-case}
	
	
	The enforcement workflow consists of five stages. First, during \emph{admission}, the Policy Interpreter converts the experiment request into a spectrum grant specifying its frequency, bandwidth, and power constraints. The Experiment Steward then deploys the experiment and verifies its configuration using calibrated spectrum measurements.
	
	Second, a controlled \emph{drift} is introduced by switching the experiment to full-power, full-bandwidth operation. Due to the spectrum inversion, the resulting emission skirt increases interference in the lower-adjacent \gls{fss} sub-band and eventually exceeds the protection budget.
	
	Third, the Spectrum Sentinel performs periodic FieldFox captures and maps the measured \gls{if} power to the virtual \gls{fss} victim reference plane. It detects the violation and correlates the observed spectrum with the experiment configuration. The lower-adjacent emission provides evidence of the inverted-skirt mechanism.
	
	Fourth, the Zone Orchestrator receives the attribution, candidate mitigations, and their projected post-action interference and disruption cost, then selects the lowest-cost candidate that meets the protection target. The selected action is a \gls{prb}-region mask that removes the \glspl{prb} mapped to the protected \gls{fss} sub-band while retaining the remaining resources. The proposed action is then submitted to the policy gate, which independently computes its expected outcome before admitting it.
	
	Finally, the Experiment Steward applies the mask through the \gls{oran} control plane, and the Spectrum Sentinel performs an independent post-action measurement to verify that the interference budget is restored. If successful, the experiment continues with the reduced allocation rather than being terminated. In contrast, the procedural baseline shuts down the experiment upon the same violation.
	
	Throughout the experiment, the reflex independently monitors the budget and can immediately apply the pre-computed \gls{prb} mask, providing a deterministic protection path independent of agent execution.
	
	\begin{table}[t]
		\caption{Experiment parameters.}
		\label{tab:params}
		\centering
		\footnotesize
		\begin{tabular}{@{}l p{4.6cm}@{}}
			\toprule
			\textbf{Parameter} & \textbf{Value} \\
			\midrule
			\acrshort{nr} numerology / \acrshort{prb}s & $\mu = 1$ ($30$\,kHz \acrshort{scs}) / $106$ \\
			DL/UL pattern & DDDDDDDSUU \\
			gNB TX-power cap $P_{\max}$ & 20 dBm \\
			B210 TX gain range & 0--89.75 dB \\
			B210 max RX input & $-15$ dBm \\
			\acrshort{if} carrier (\acrshort{fr1}) & $3.58$\,GHz \\
			\acrshort{fr3} \acrshort{lo} / injection & $13.58$\,GHz / high-side  \\
			\acrshort{fr3} carrier &  $10$\,GHz \\
			RF interconnect & conducted \acrshort{if} chains (attenuators, splitters, BPF/LPF); \gls{ota} ($2$\,m Tx/Rx) \\
			Control interval $\Delta t$ & $100$\,ms \\
			Protection criterion & \acrshort{itur} S.1432 \acrshort{in} $-10$\,dB \\
			Victim realization & virtual (FieldFox + S.1428/S.1432) \\
			Victim noise / ref. bandwidth & $T_{\mathrm{sys}} = 150$\,K / $B = 36$\,MHz \\
			Traffic & UDP DL iperf3 \\
			Agent runtime  & Claude Sonnet 5 \\
			Trials per condition $N_{\sf trial}$ & $10$ \\
			\bottomrule
		\end{tabular}
		\vspace{-10pt}
	\end{table}

    \vspace{-2mm}
	\section{Numerical Results}
	\label{sec:results}
    \vspace{-1mm}
	
	We evaluate \ardz{} over $N_{\sf trial}\!=\!10$ drift trials and compare it with a procedural shutdown baseline. All enforcement stages operate on the hardware-in-the-loop path: FieldFox measurements provide the interference estimate, which is referred to the virtual victim plane using the calibration of Sec.~\ref{sec:setup}. Post-actuation verification uses an independent capture. All timestamps are wall-clock measurements from a single host. The key experimental parameters are summarized in Table~\ref{tab:params}.
	
	\vspace{-1mm}
	\subsection{Detection-to-Mitigation Timeline}
	\label{sec:res-latency}
	\vspace{-1mm}
	
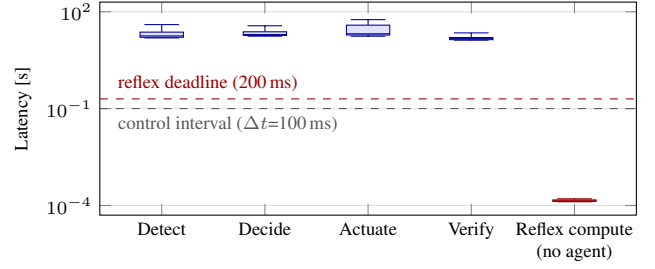
\begin{figure}[t]
		\centering
	\begingroup
	\footnotesize
	\begin{tikzpicture}
		\begin{axis}[
			width=0.98\linewidth, height=4.4cm,
			ymode=log, ymin=5.01187e-05, ymax=199.526,
			ylabel={Latency [s]},
			ylabel near ticks,
			xmin=0.4, xmax=5.6,
			xtick={1,2,3,4,5},
			xticklabels={Detect, Decide, Actuate, Verify, {Reflex compute\\(no agent)}},
			xticklabel style={align=center, font=\scriptsize},
			boxplot/draw direction=y,
			boxplot={box extend=0.42},
			ymajorgrids, grid style={gray!25},
			tick label style={font=\scriptsize},
			label style={font=\scriptsize},
			clip=false,
			]
			\addplot[fill=blue!12, draw=blue!60!black, solid, boxplot prepared={
				lower whisker=15.479000, lower quartile=16.222000, median=17.876000,
				upper quartile=23.397000, upper whisker=40.052000}] coordinates {};
			\addplot[fill=blue!12, draw=blue!60!black, solid, boxplot prepared={
				lower whisker=17.575500, lower quartile=18.534000, median=19.637000,
				upper quartile=23.993500, upper whisker=36.845000}] coordinates {};
			\addplot[fill=blue!12, draw=blue!60!black, solid, boxplot prepared={
				lower whisker=17.488500, lower quartile=19.084000, median=20.798000,
				upper quartile=38.486000, upper whisker=56.979500}] coordinates {};
			\addplot[fill=blue!12, draw=blue!60!black, solid, boxplot prepared={
				lower whisker=13.212500, lower quartile=13.678000, median=14.990000,
				upper quartile=15.929000, upper whisker=22.171000}] coordinates {};
			\addplot[fill=red!12, draw=red!60!black, solid, boxplot prepared={
				lower whisker=0.000130, lower quartile=0.000133, median=0.000140,
				upper quartile=0.000149, upper whisker=0.000161}] coordinates {};
			\draw[dashed, gray!70!black]
			(axis cs:0.4,0.1) -- (axis cs:5.6,0.1)
			node[pos=0.02, below, font=\scriptsize, gray!70!black, anchor=north west]
			{control interval ($\Delta t$=100\,ms)};
			\draw[dashed, red!60!black]
			(axis cs:0.4,0.2) -- (axis cs:5.6,0.2)
			node[pos=0.02, above, font=\scriptsize, red!60!black, anchor=south west]
			{reflex deadline (200\,ms)};
		\end{axis}
	\end{tikzpicture}
	\endgroup
		\caption{Latency decomposition of the agentic enforcement loop over $N_{\rm trial}=10$ trials.}
		\label{fig:timeline}
		\vspace{-4mm}
	\end{figure}

	Figure~\ref{fig:timeline} compares the latency of the agentic enforcement path with the agent-independent reflex. 
    The reflex deadline of $200$\,ms represents the design sense-to-actuate budget of two $100$\,ms control intervals.
    With the agent, the enforcement loop includes detection, decision, actuation, and verification. The median end-to-end latency is $73.74$\,s, obtained as the sum of the four per-stage medians plotted in Fig.~\ref{fig:timeline}.
	In contrast, the reflex decision itself requires only $0.14$\,ms of computation on average because it applies a pre-computed protective action without \gls{llm}-based reasoning or agent-mediated tool execution. The end-to-end protection response remains bounded by the $100$\,ms sensing/control interval and the latency of the sensing path.
	This difference highlights the complementary roles of the two paths: the reflex provides fast, deterministic protection, whereas the agentic path performs higher-level diagnosis, mitigation selection, and post-action verification.
	
	\begin{figure}[t]
		\centering
\begingroup
\footnotesize
\begin{tikzpicture}
	\begin{axis}[
		width=0.98\linewidth, height=4.6cm,
		xlabel={Time [s]}, ylabel={Margin to budget [dB]},
		xmin=0, xmax=79.69, ymin=-10.29, ymax=10.0,
		ymajorgrids, grid style={gray!25},
		tick label style={font=\scriptsize},
		label style={font=\scriptsize},
		ylabel near ticks, xlabel near ticks,
		xtick pos=left, ytick pos=left,
		legend style={font=\scriptsize, at={(0.98,0.12)}, anchor=south east, align = left,
			draw=gray!50, fill=white, fill opacity=0.85, text opacity=1},
		clip=false,
		]
		\fill[red!8] (axis cs:0,-10.29) rectangle (axis cs:79.69,0);
		\draw[black, thick] (axis cs:0,0) -- (axis cs:79.69,0)
		node[pos=0.05, below right, font=\scriptsize, black]
		{$I^{\max}_{\rm FSS}$ (I/N $\le$ -10\,dB, ITU-R S.1432)};
		
		\addplot+[blue!50!black, thick, mark=none] coordinates {
			(0.0000,3.892) (5.0000,3.892) (5.0000,-8.294) (61.5830,-8.294) (61.5830,3.048) (79.6890,3.048)
		};
		\addlegendentry{A-RDZ: masked, exp. continues (cost 0.24)}
		
		\addplot+[black, dashed, thick, mark=none] coordinates {
			(0.0000,3.892) (5.0000,3.892) (5.0000,-8.294) (5.0001,-8.294) (5.0001,6.000) (79.6890,6.000)
		};
		\addlegendentry{baseline: shutdown (cost 1.00)}
		
		\draw[gray!60] (axis cs:5.000,0) -- (axis cs:5.000,9.0);
		\node[font=\scriptsize, anchor=south, align=center] at (axis cs:5.000,10) {drift\\[-2pt](full buffer)};
		\draw[gray!60] (axis cs:21.660,0) -- (axis cs:21.660,9.0);
		\node[font=\scriptsize, anchor=south, align=center] at (axis cs:21.660,10) {detect};
		\draw[gray!60] (axis cs:61.583,0) -- (axis cs:61.583,9.0);
		\node[font=\scriptsize, anchor=south, align=center] at (axis cs:61.583,10) {mask\\[-2pt]applied};
		\draw[gray!60] (axis cs:74.689,0) -- (axis cs:74.689,9.0);
		\node[font=\scriptsize, anchor=south, align=center] at (axis cs:74.689,10) {verified};
	\end{axis}
\end{tikzpicture}
\endgroup
		\caption{Interference margin, estimated at the virtual \acrshort{fss} victim reference plane, during one representative trial.}
		\label{fig:trace}
		\vspace{-4mm}
	\end{figure}
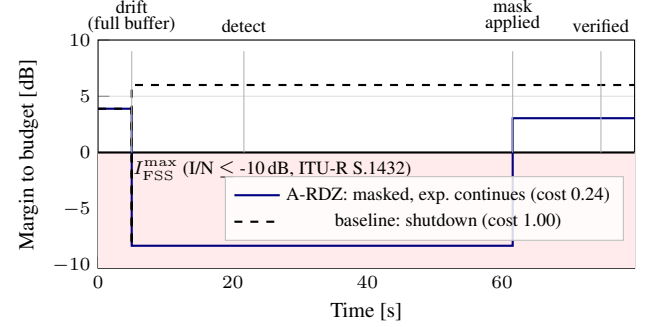
	
	\vspace{-2mm}
	\subsection{Protection Compliance and Experiment Utility}
	\label{sec:res-protection}
	\vspace{-1mm}
	
	We define the instantaneous margin relative to the protection budget as 
	$ \text{Margin}(t) \!=\! 10 \log_{10}(I^{\max}_{\mathrm{FSS}} / I(t))
	$\,[dB],
	so that a positive margin denotes compliance and a negative margin denotes a violation of $I_{\mathrm{FSS}}^{\max}$.
	Figure~\ref{fig:trace} shows a representative drift episode. 
    At the detected violation, $I/I_{\mathrm{FSS}}^{\max}=6.75$, corresponding to a margin of $-8.29$\,dB. The agent selects a PRB-region mask, reducing the interference to $I/I_{\mathrm{FSS}}^{\max}=0.496$ and restoring a positive margin of $3.05$\,dB.
	
	We also define the utility retained by a mitigated allocation $\mathbf{p}$ as $\eta = \sum_{n} e\big(m_n(p_n)\big) \,/\, \sum_{n} e\big(m_n(p_n^{\mathrm{grant}})\big)$, the ratio of achievable downlink rate to that of the granted operating point, where $e(\cdot)$ is the TS~38.214 spectral-efficiency mapping~\cite{ts38214} and $m_n$ denotes the \gls{mcs} index selected for \gls{prb} $n$ under allocation $p_n$. The disruption cost is $C \! = \! 1 \!-\! \eta$, so a full shutdown gives $C \! = \! 1$.
	Across all 10 trials, the violation is successfully detected and mitigated, and the experiment retains $76.04$\% of its utility ($\eta\!=\!0.76$).
	In comparison, the shutdown baseline terminates the experiment, corresponding to a normalized cost of $C\!=\!1.00$ versus $C\!=\!0.24$ for \ardz{}.

	\section{Related Work}
	\label{sec:related}
	
	\textbf{RDZ concept and prototypes.} The \gls{rdz} concept, its stakeholder analysis, and its policy/experiment/spectrum decomposition were articulated in~\cite{Zheleva2023RDZ}. 
	Prototype zone management systems include Powder-RDZ and its OpenZMS zone management system~\cite{PowderRDZ2024,OpenZMS2025}, the COSMOS ZMS~\cite{CosmosZMS2026}, and reactive spectrum-sharing extensions on POWDER~\cite{Sarbhai2025ASTRA}. FlexRDZ~\cite{FlexRDZ2023} is closest in spirit to the autonomy, managing mobile transmitters with hierarchical task networks and digital-twin prediction; its planning, however, remains procedural over a fixed task vocabulary, whereas the \ardz{} reasons from heterogeneous evidence while enforcing policy constraints.
	
	\textbf{Agentic AI for wireless.} \genesis{}~\cite{GENESIS} introduced multi-agent autonomous R\&D on production \gls{oran} testbeds and is the substrate we extend; AgentRAN~\cite{AgentRAN} proposes an agentic architecture for autonomous control of open 6G networks, and ALLSTaR~\cite{ALLSTaR} uses \glspl{llm} to synthesize and test \gls{mac} schedulers from intent.
	To our knowledge, no prior work has applied agentic systems to the \emph{operation of a spectrum-sharing zone}, where the object of control is not a network's own performance but the protection relationship between an experiment and an incumbent. 
	
	\textbf{FR3 coexistence.} Upper mid-band \gls{nr}--\gls{fss} coexistence has been studied predominantly through normative emission masks and statistical propagation models~\cite{ITU_M2101,Lim_TerSat_Coex}, with \gls{fr3} channel measurements now emerging from real hardware~\cite{NYU_FR3_urban,NYU_FR3_indoor}.  
	Our work complements these efforts by demonstrating agentic zone management over a hardware-in-the-loop FR3 coexistence experiment.
	
	\section{Conclusions}
	\label{sec:conclusion}
	
	We introduced \ardz{}, an agentic realization of the \gls{rdz} architecture in which LLM-based agents perform spectrum, experiment, and policy management while deterministic gates and \gls{nrt} reflexes preserve incumbent protection. We demonstrated the architecture end-to-end on a hardware-in-the-loop \gls{fr3} \gls{oran} testbed with a 5G NR experiment and an emulated \gls{fss} incumbent.
	
	The results show that agentic operation can detect and mitigate unexpected interference while retaining experiment utility, but its non-deterministic latency motivates separating adaptive reasoning from safety-critical protection. Future work will address multi-experiment coordination, boundary sensing, formal verification of the agent--gate interface, and larger-scale \gls{rdz} deployment.
	
	\vspace{-2mm}
	\bibliographystyle{IEEEtran}
	\bibliography{ref_s26_fr3}
	
\end{document}